\documentclass[a4paper]{jpconf}
\usepackage{graphicx}

\usepackage{citesort}

\newcommand{\softsh}[1]{\texttt{#1}}
\newcommand{\soft}[1]{\texttt{#1 }}

\newcommand{\lisepp}{\soft{LISE$^{++}$}}
\newcommand{\liseppsh}{\softsh{LISE$^{++}$}}

\newcommand{\epaxtwo}{\soft{EPAX 2.15}}
\newcommand{\epaxthree}{\soft{EPAX~3.1}}

\begin{document}


\title{Production cross sections of neutron rich isotopes from a $^{82}$Se beam}

\author{
O.~B.~Tarasov$^1$,
D.~J.~Morrissey$^{1,2}$,
A.~M.~Amthor$^3$,
L.~Bandura$^3$,
T.~Baumann$^1$,
D.~Bazin$^1$,
J.~S.~Berryman$^1$,
G.~Chubarian$^4$,
N.~Fukuda$^5$,
A.~Gade$^{1,6}$,
T.~N.~Ginter$^{1}$,
M.~Hausmann$^{3}$,
N.~Inabe$^5$,
T.~Kubo$^5$,
J.~Pereira$^{1}$,
M.~Portillo$^{3}$,
B.~M.~Sherrill$^{1,6}$,
A.~Stolz$^{1,6}$,
C.~Sumithrarachchi$^1$,
M.~Thoennessen$^{1,6}$,
D.~Weisshaar$^1$
}

\address{$^1$ National Superconducting Cyclotron Laboratory, Michigan State University, East Lansing, MI 48824, USA }
\address{$^2$ Department of Chemistry, Michigan State University, East Lansing, MI 48824, USA}
\address{$^3$ Facility for Rare Isotope Beams, Michigan State University, East Lansing, MI 48824, USA}
\address{$^4$ Cyclotron Institute, Texas A\&M University, College Station, TX 77843, USA}
\address{$^5$ RIKEN Nishina Center, RIKEN, Wako-shi, Saitama 351-0198, Japan}
\address{$^6$ Dep. of Physics and Astronomy, Michigan State University, East Lansing, MI 48824, USA}

\ead{tarasov@nscl.msu.edu}

\begin{abstract}

Production cross sections for neutron-rich nuclei from the
fragmentation of a $^{82}$Se beam at 139~MeV/u were measured. The
longitudinal momentum distributions of 122 neutron-rich isotopes of
elements  \protect{$11\le Z\le 32$} were determined by varying the target thickness. Production
cross sections with beryllium and tungsten targets were determined
for a large number of nuclei including several isotopes first observed in
this work. These are the most neutron-rich nuclides of the elements
\protect{$22\le Z\le 25$} ($^{64}$Ti, $^{67}$V, $^{69}$Cr,
$^{72}$Mn). One event was registered consistent with $^{70}$Cr, and another one with
$^{75}$Fe. A one-body $Q_{g}$ systematics is used to
describe the production cross sections based on thermal evaporation
from excited prefragments. The current results confirm those of our previous experiment with a $^{76}$Ge beam:
 enhanced production cross sections for neutron-rich fragments near $Z=20$.
\end{abstract}


\section{Introduction\label{Intro}}

The discovery of new nuclei in the proximity of the neutron dripline
provides a test for nuclear mass models, and hence for the
understanding of the nuclear force and the creation of elements.
Once neutron-rich nuclei are observed, and their cross sections for  formation
are understood,  investigations to study
the nuclei themselves, such as decay spectroscopy, can be planned.
Therefore,  obtaining  production rates for the most exotic
nuclei continues to be an important part of the experimental program
at existing and future rare-isotope facilities.

A number of production mechanisms have been used to produce
neutron-rich isotopes for  \protect{$20\le Z\le 28$} ~\cite{OT-PRC09}. But in the last years two reaction mechanisms
were the most effective at producing nuclei in this region:
\begin{itemize}
  \item projectile fragmentation -- an experiment with a $^{76}$Ge (132 MeV/u) beam produced 15 new isotopes of  \protect{$17\le Z\le 25$}~\cite{OT-PRL09},
  \item in-flight fission with light targets (Abrasion-Fission) -- an experiment with a $^{238}$U beam~\cite{Ohn-JPSJ10} produced  a large number of isotopes of  \protect{$25\le Z\le 48$} using a Be-target, and several new isotopes with \protect{$46\le Z\le 56$} by Coulomb fission on a heavy target.
\end{itemize}

\begin{figure*}
\begin{center}
\includegraphics[width=0.95\textwidth]{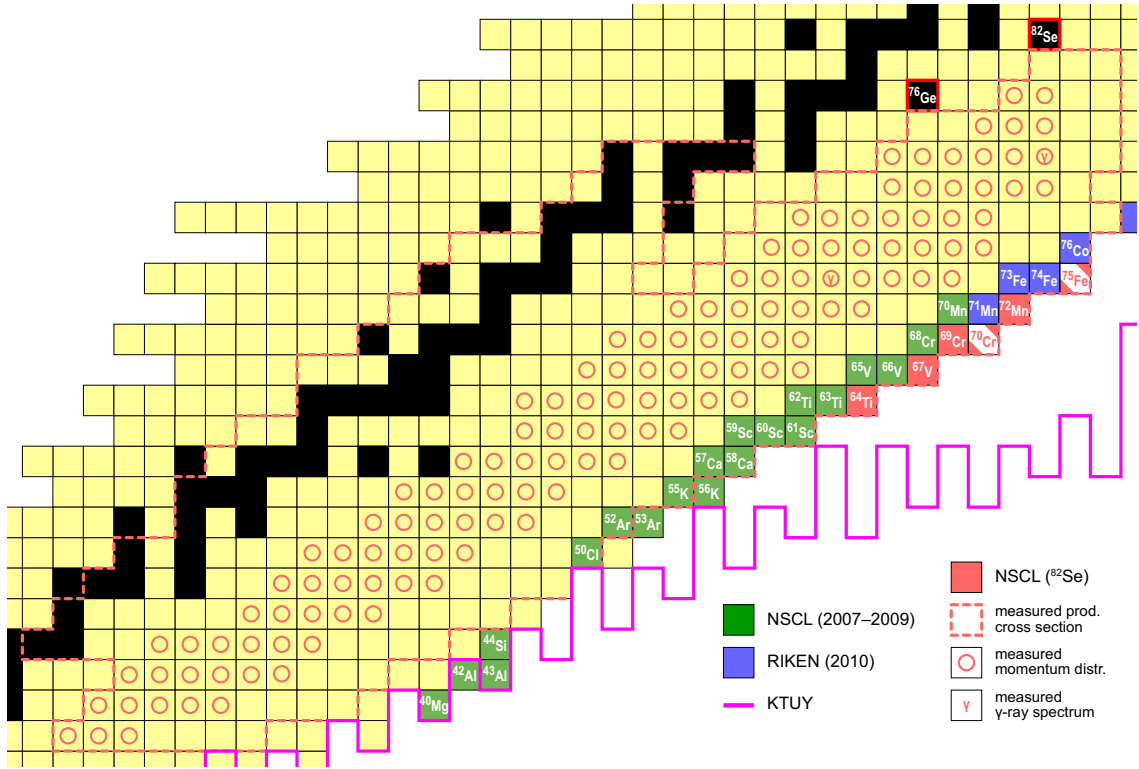}
\caption{(Color online) The region of the nuclear chart investigated
in the   present work. The solid line shows the limit of bound nuclei from the
KTUY mass model~\cite{KTUY-PTP05}. The new isotopes observed for
the first time in the present work are marked by red
squares.\label{chart}}
\end{center}
\end{figure*}

Progress in the production of neutron-rich isotopes was made possible
by the increase of primary beam intensities, new beam development at the National
Superconducting Cyclotron Laboratory (NSCL) at Michigan State
University and advances in experimental techniques \cite{TB-N07}.
Indeed, recent measurements
at the NSCL~\cite{OT-PRC07,TB-N07,PFM-BAPS08,OT-PRC09} have demonstrated that
the fragmentation of $^{48}$Ca and $^{76}$Ge beams can be used to
produce new isotopes in the proximity of the neutron dripline.
Continuing this work, we report here the next step with a newly developed $^{82}$Se beam towards the
fundamental goal of defining the absolute mass limit for chemical
elements in the region of calcium. In the present measurement, four neutron-rich
isotopes with \protect{$42\le N\le 47$} were identified for the
first time (see Fig.\ref{chart}), one event was registered consistent with $^{70}$Cr$_{46}$, and another one with
$^{75}$Fe$_{49}$.

One of the first indications of significant changes in the structure of neutron rich nuclei was the discovery of enhanced nuclear binding of heavy sodium isotopes~ \cite{CT-PRC75}. This is now understood to result from significant contributions of {\it fp} shell intruder orbitals to the ground-state configurations of these isotopes~ \cite{XC-NPA75,EKW-PRC90}. Low-lying 2$^+$ states and quadrupole collectivity have been reported in neutron-rich even-even Ne and Mg isotopes around $N=20$, see for example Refs.~\cite{DGM-NPA84,MO-PLB95,YO-PLB01,YA-PLB03,CH-PRC05}.
This region around $^{31}$Na, where the neutron {\it fp} shell contributes significantly to the ground-state structure, is now known as the ``Island of Inversion''. Similarly, there is mounting evidence for an onset of deformation around neutron number $N=40$ in Fe and Cr nuclei. In even-even Fe and Cr nuclei, for example, this evidence is based on the energies of low-lying states~\cite{HA-PRL99,SO-EPJA03,AG-PRC08,AO-PRL09,GA-PRC10}, transition strengths~\cite{RO-PRL11}, and deformation length ~\cite{AO-PRL09}. Neutron $g_{9/2}$ and $d_{5/3}$ configurations from above the $N=40$ shell gap are proposed to descend and dominate the low-lying configurations similar to the $N=20$ Island of Inversion~\cite{BAB-PPNP01,LE-PRC10}.

In our previous cross section measurements in the region around $^{62}$Ti ($^{76}$Ge primary beam)~\cite{OT-PRL09} we observed  a systematic smooth variation of the production cross sections that might point to nuclear structure effects, for example, an onset of collectivity, that are not included in global mass models that form the basis of systematics. The present work, since using a different primary beam, provides an independent check of this interpretation.

\section{Experiment\label{secExpt}}
\subsection{Setup\label{secSetup}}

A newly developed 139 MeV/u $^{82}$Se beam with an intensity of 35~pnA, accelerated by the coupled cyclotrons at
the NSCL, was fragmented on a series of beryllium targets and a
tungsten target, each placed at the object position of the A1900
fragment separator~\cite{DJM-NIMA03}. In this work we used an identical configuration as in our previous experiment with a $^{76}$Ge beam ~\cite{OT-PRC09},
where the combination of the A1900 fragment separator with
the S800 analysis beam line~\cite{DB-NIMB03} formed a two-stage
separator system, that allowed a high
degree of rejection of unwanted reaction products. At the end of the S800 analysis beam line, the
particles of interest were stopped in a telescope of eight silicon
PIN diodes (50$\times$50~mm$^2$) with a total thickness of 8.0~mm.  A 50~mm thick plastic scintillator
positioned behind the Si-telescope served as a veto detector against
reactions in the Si-telescope and provided a measurement of the
residual energy of lighter ions that were not stopped in the
Si-telescope. A position sensitive parallel plate avalanche counter
(PPAC) was located in front of the Si-telescope.
All experimental details and a sketch of the experimental setup can be found in Ref.~\cite{OT-PRC09}.
In this paper, we describe the details of our  experimental approach and discuss the
results.


\subsection{Experimental runs\label{secPlanning}}

The present experiment consisted of four  parts. Except for the last part, the present experiment planning 
was similar to the previous  $^{76}$Ge experiment~\cite{OT-PRC09}.
During all runs, the magnetic rigidity of
the last two dipoles of the analysis line was kept constant at a nominal value of 4.3~Tm while the
production target thickness was varied to map the fragment momentum
distributions. This approach  greatly simplified the particle
identification during the scans of the parallel momentum
distributions.

The   momentum acceptance of the A1900 fragment separator was
restricted to $\Delta  p/p = 0.1\%$ (first four runs with thin targets), and $\Delta  p/p = 0.2\%$
 (other  targets) for the measurement of differential momentum
distributions in the first part of the
experiment. The use of different beryllium target thicknesses
(9.7, 68, 138, 230, 314, 413, 513~mg/cm$^2$) allowed coverage of the fragment momentum distributions necessary to extract
production cross sections and also resulted in more isotopes in the
particle identification spectrum.

 For the second part of the experiment, a Kapton wedge
with a thickness of 20.0 mg/cm$^2$ was used at the dispersive image of
the A1900 to reject less exotic fragments with  a 10~mm aperture in
the focal plane while the separator was set for $^{67}$Fe and $^{78}$Zn ions. The goal
of this setting was to confirm the particle identification by isomer
tagging as described in
Ref.~\cite{RG-PLB95} with $^{67m}$Fe~($E_{\gamma}=367$~keV, $T_{1/2}=43$~$\mu$s)
and $^{78m}$Zn~($E_{\gamma}=730, 890, 908$~keV, $T_{1/2}=0.32$~$\mu$s).

In the third part of the experiment, dedicated to the  search  for
 new isotopes, five settings were  used to cover the most
neutron-rich isotopes  with \protect{$20\le Z\le 27$}, as it was
impossible to find a single target thickness and magnetic rigidity
to produce all of the fragments of interest. Each setting was
characterized by a fragment for which the separator was tuned. A
search for the most exotic nuclei in each setting was carried out
with Be and W targets. The settings were centered on $^{60}$Ca,
$^{68}$V and $^{74,75}$Fe respectively, based on
\liseppsh~\cite{OT-NIMB08} calculations using the parameterizations
of the momentum distributions obtained in the first part of the
experiment (see Section~\ref{secMomentum}). The momentum acceptance
of the A1900 was set to the maximum of $\Delta p/p = 5.0\%$ for
these production runs. It is noteworthy that the momentum acceptance of the S800 beamline  is about 4\%
according to \lisepp Monte Carlo simulations based on new extended configurations with 5-order optics.
These calculations have been taken into account for the cross section analysis.

The fourth part of the experiment has been devoted to two short runs to measure the yield of more stable isotopes
by centering on $^{45,48}$Ca.

\begin{figure}
\centering
\includegraphics[width=0.65\columnwidth]{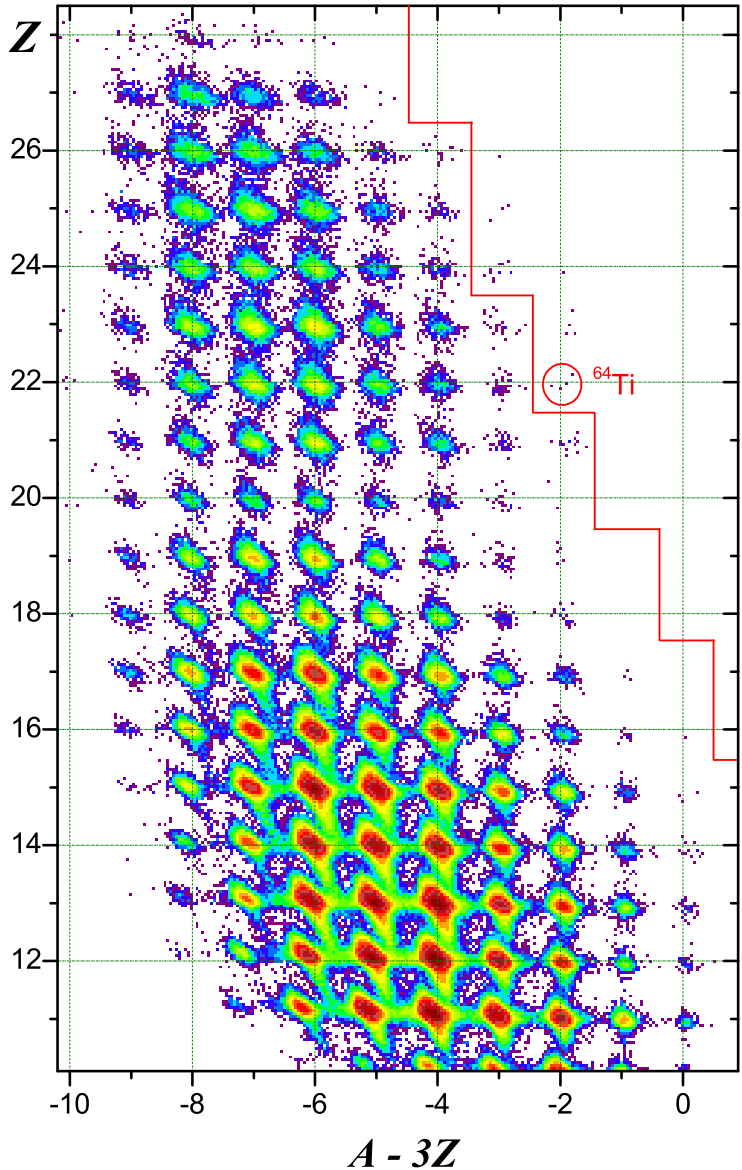}%
\caption{(Color online) Particle identification plot showing the
measured atomic number, $Z$, versus the calculated function $A-3Z$
for the nuclei observed in production runs of this work. See text for details. The limit of previously observed
nuclei is shown by the solid red line as well as the locations of
a reference nucleus ($^{64}$Ti). \label{Fig_pid}}
\end{figure}

\section{Analysis of experimental data \label{secAnalysis}}

\begin{figure*}
\begin{center}
\includegraphics[width=0.9\textwidth]{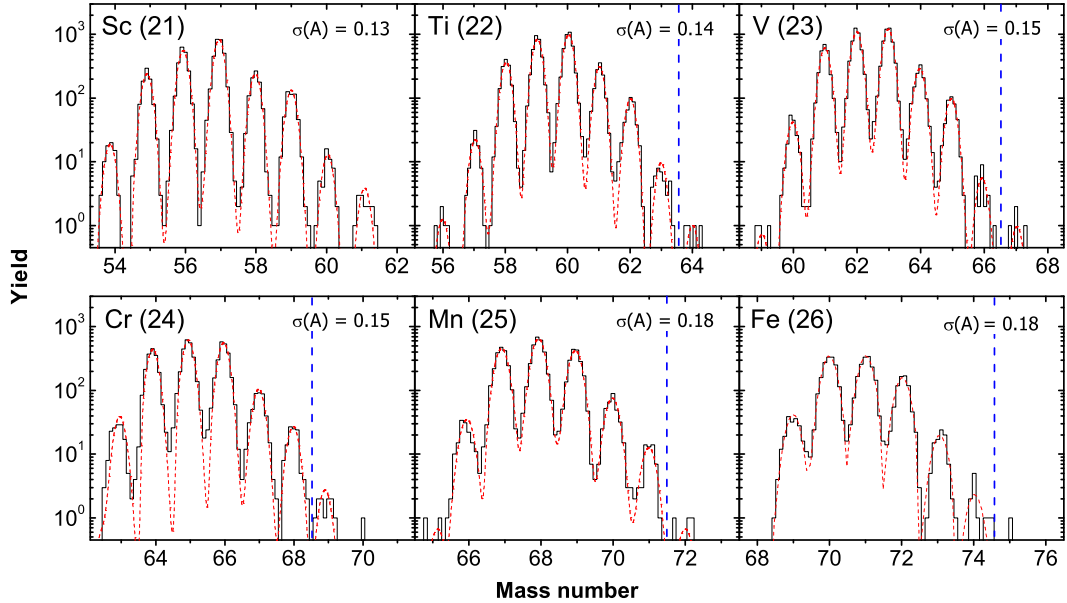}%
\caption{(Color online) Mass spectra of the elements \protect{$21\le Z\le 26$}.
 All particles that were stopped in the Si-telescope
during the production runs were analyzed. The limits of previously
observed nuclei are shown by the vertical dashed lines.
Standard deviations produced in multi-peak fits with Gaussian
distributions at constant width (dashed curves) are shown in the
figures for each element.
\label{Fig_Mass}}
\end{center}
\end{figure*}


The particle identification matrix of fully-stripped reaction products observed in the production runs
 is shown in Fig.~\ref{Fig_pid}. The range of fragments is shown as a function of the measured
atomic number, $Z$,  versus the calculated quantity $A-3Z$.
The identification of the individual isotopes in Fig.~\ref{Fig_pid}
was confirmed via isomer tagging using the known isomeric decays in
$^{67}$Fe and $^{78}$Zn. The
standard deviations of ionic ($q$) and elemental ($Z$) spectra  were found to be similar to those in the previous experiment,
therefore the probabilities of one event being
misidentified as a neighboring charge state or element were as small as before.
The details of the calculation of the particle identification are given
in the appendix of the previous work~\cite{OT-PRC09}.

\subsection{Search for new isotopes\label{secNewIsotopes}}
The mass spectra for the isotopic chains from scandium to iron
measured during the production runs are shown in
Fig.~\ref{Fig_Mass}. Only nuclei that stopped in the Si telescope are included in this analysis. The observed fragments include
several new isotopes that are the most neutron-rich nuclides yet observed
of elements \protect{{$22\le Z\le 25$} ($^{64}$Ti, $^{67}$V, $^{69}$Cr,
$^{72}$Mn)}. One event was found to be consistent with $^{70}$Cr, and another one with
$^{75}$Fe.  The new neutron-rich nuclei
observed in this work are those events to the right of the solid
line in Fig.~\ref{Fig_pid} and to the right of the vertical dashed
lines in Fig.~\ref{Fig_Mass}.

\subsection{Parallel momentum distributions\label{secMomentum}}

The prediction of the momentum distributions of
residues is important when searching for new isotopes  in order to set the
fragment separator at the maximum production rate. Also, the
accurate prediction of the momentum distributions allows a precise estimate
of the transmission and efficient rejection of strong contaminants.
In this experiment the ``target scanning" approach~\cite{OT-NIMA09} developed in the previous experiment
was used to obtain parameters for the neutron-rich isotope momentum distribution models such as ~\cite{AG-PLB74,DJM-PRC89}.
This method is particularly well suited to survey
neutron-rich nuclei since the less exotic nuclei are produced with
the highest yields and their momentum distributions can be measured
with the thin targets.

The data analysis of this approach has been updated, and a detailed explanation is in preparation~\cite{OT-NIMA12}.
Important improvements include: first, that the most probable velocity for a fragment is not
that at the center of the target if the yield is sharply rising  or falling with momentum,
and second, asymmetric Gaussian distributions have been used where the asymmetry coefficients
have been taken from the convolution model implemented in the \lisepp code~\cite{OT-NIMB08}. Note that, at the energy of these experiments,
 the shape of the fragment momentum distribution is slightly asymmetric with a
low-energy exponential tail stemming from dissipative
processes~\cite{OT-NPA04}. Seven targets were used to measure the momentum distributions.
The momentum distributions for 122 isotopes were derived and integrated to deduce the production cross sections.
A survey of all of the fitted results showed that neutron-rich fragments were produced with significantly higher velocities than the momentum distribution models~\cite{DJM-PRC89,BOR-ZPA83} predict, and this result is similar to our previous measurements~\cite{OT-NIMA09}.

\section{Results and Discussion\label{secRes}}


\subsection{Production cross section\label{secCS}}

\begin{figure*}
\includegraphics[width=0.9\textwidth]{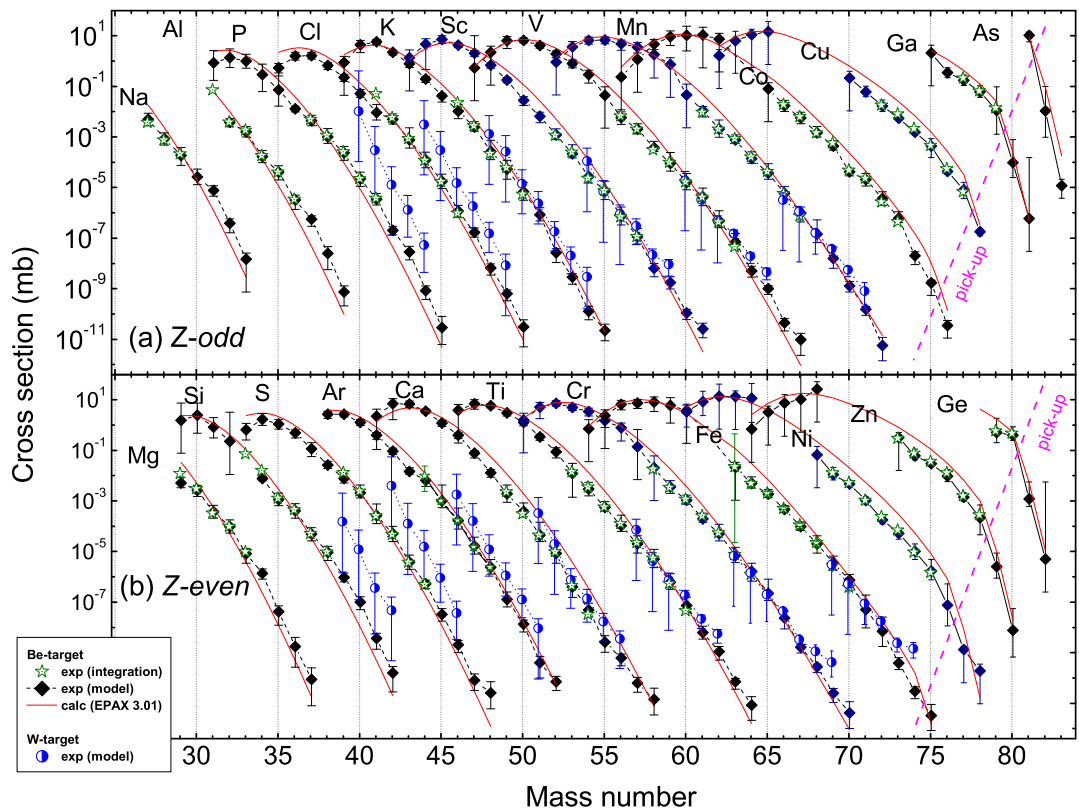}%
\caption{(Color online) Inclusive production cross sections for
fragments from the reaction of $^{82}$Se with beryllium and tungsten
targets shown as a function of mass number. The cross sections with
the beryllium targets derived by momentum distribution integration
are shown by stars, those normalized with \liseppsh\ transmission
calculations are indicated by solid diamonds The cross sections
obtained with the tungsten target were normalized with \liseppsh\
transmission calculations.  The
red solid lines show the predictions of the \epaxthree systematics~\cite{KS-EPAX3} for
beryllium.
The two magenta dashed lines separate
nuclei that require neutron pickup in the production mechanism.
\label{Fig_CrossSection}}
\end{figure*}

The inclusive production cross sections for the observed fragments
were calculated by correcting the measured yields for the finite
momentum and angular acceptances of the separator system. A
total of 122 cross sections with beryllium were obtained
from the Gaussian fits to the longitudinal momentum
distributions; these nuclei are indicated by stars in
Fig.~\ref{Fig_CrossSection}. The cross sections for all of the
remaining fragments with incompletely measured longitudinal momentum
distributions were obtained with estimated transmission corrections as has been done in our previous work~\cite{OT-PRC09}.
The angular and momentum transmissions were calculated for each
isotope in each setting using a model of the momentum distribution
with smoothly varying parameters extracted from the measured
parallel momentum distributions.

The cross sections obtained for all of the fragments observed in
this experiment are shown in Fig.~\ref{Fig_CrossSection} along with
the predictions of the recent \epaxthree parameterization~\cite{KS-EPAX3}.  For those isotopes that relied on
transmission calculations, the weighted mean of all measured yields
was used to obtain the ``model-based'' cross section (shown by solid
diamonds in Fig.~\ref{Fig_CrossSection}). The uncertainties in these
cases include the statistical, the systematic and the transmission
uncertainties. For more details see ref.~\cite{OT-NIMA09}.
As can be seen in Fig.~\ref{Fig_CrossSection}, the
model-based cross sections are in good agreement with those produced
by integrating the measured longitudinal momentum distributions.
It is important to note that the  predictions
of the  recent \epaxthree parameterization for reactions with beryllium, shown
by the solid lines in Fig.~\ref{Fig_CrossSection}, reproduces the
measured cross sections for isotopes much better than previous \epaxtwo predictions~\cite{KS-PRC00}.


\begin{figure*}
\includegraphics[width=1.0\textwidth]{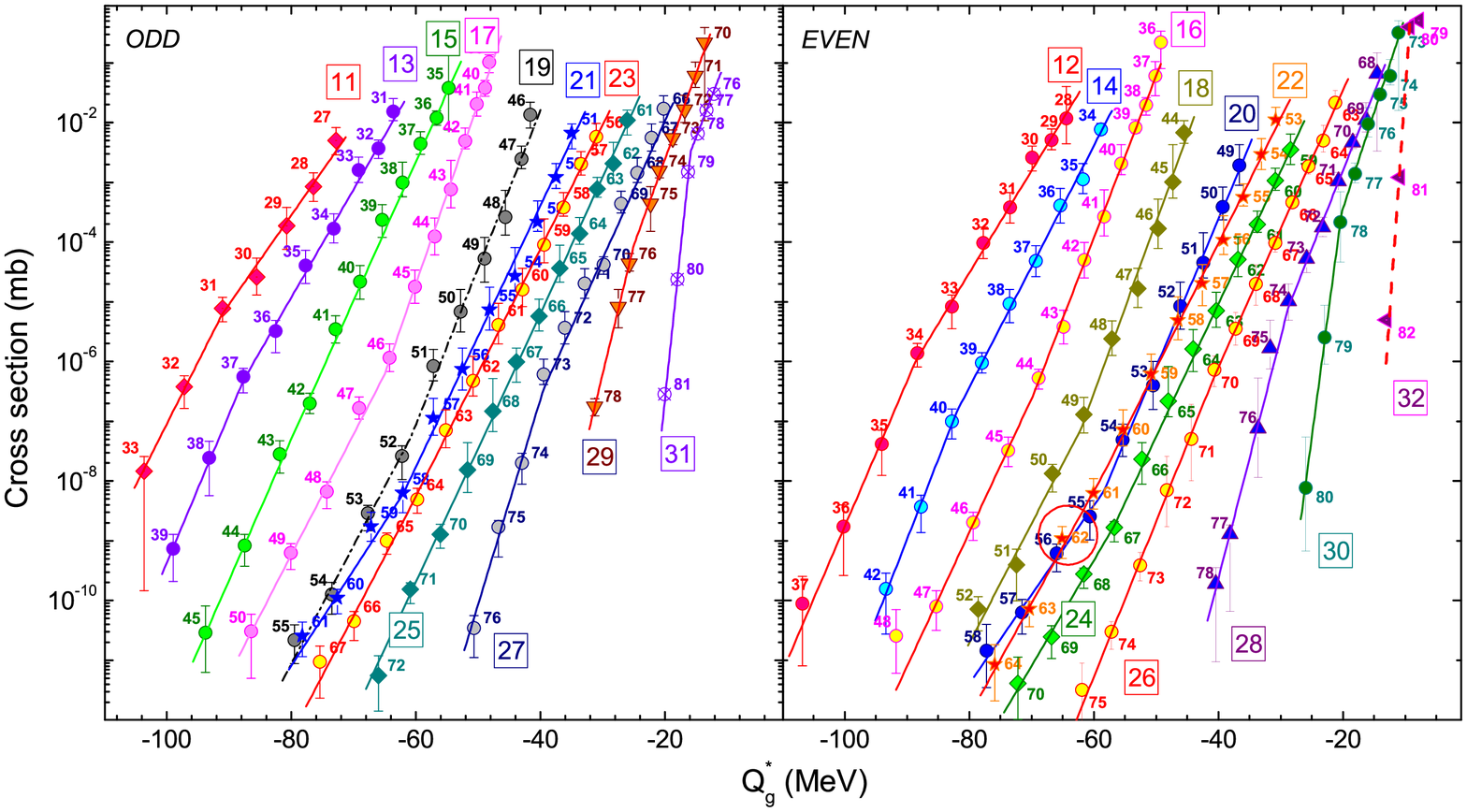}%
\caption{(Color online) Cross sections for the production of
neutron-rich nuclei with odd (left plot) and even (right plot)
atomic numbers, with a beryllium target. See text for explanation of
$Q_g^*$ and the lines. The cross section for $^{62}$Ti at the center
of the proposed new island of
inversion~\protect{\cite{BAB-PPNP01}} are circled. \\
\label{Fig_Qg}}
\end{figure*}


\subsection{Q$_g$ systematics\label{secQg}}

The production cross sections for the most neutron-rich projectile
fragments have been  previously shown to have an exponential
dependance on $Q_g$ (the difference in mass-excess of the beam
particle and the observed fragment)~\cite{OT-PRC07,OT-PRL09}.
To test this behavior, the cross sections for each isotopic chain were fitted with the simple expression:
\begin{equation}\label{Eq_Qg}
 \sigma(Z,A) = f(Z)\exp{(Q^*_{g}/T)},
\end{equation}
where  $T$ is an effective temperature  or inverse  slope.
In this work neutron odd-even corrections have been applied for $Q_g$ of neutron-odd isotopes,
that do not change slopes of lines, but smooth the data and significantly decreases the  $\chi^2$-value.
This correction has a large effect on the stable isotopes,
and practically no influence for very exotic nuclei with weakly bound neutrons.

\begin{figure}
\begin{minipage}{18pc}
\includegraphics[width=18.5pc]{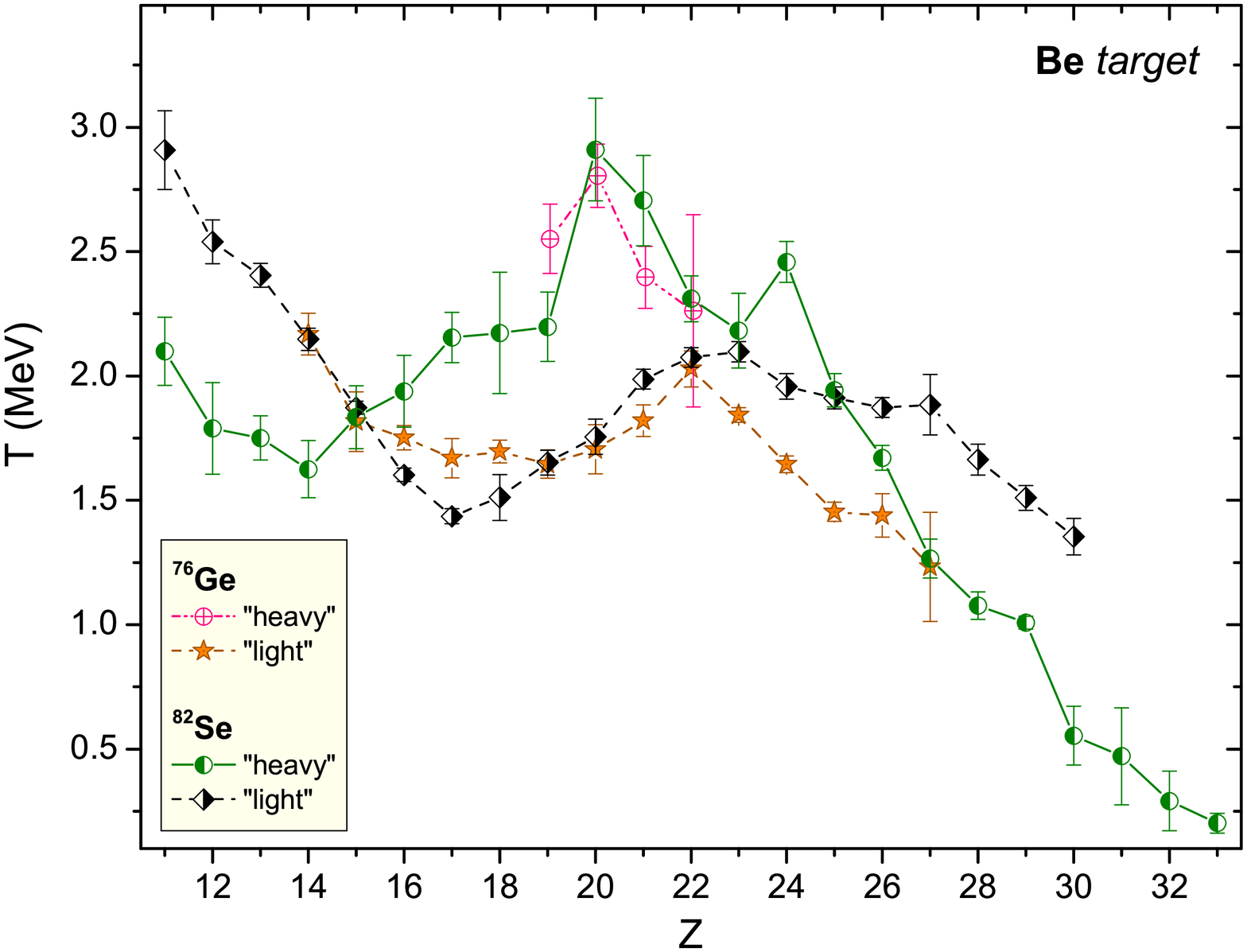}
\caption{\label{Fig_TemperatureBe}(Color online) Values of the inverse slope parameter, $T$,
from the best fit of Eq.~\ref{Eq_Qg} to the experimental cross
sections in Fig.~\ref{Fig_Qg} shown as a function of atomic number, where
half-filled green circles for heaviest isotope and half-filled black diamonds for light isotopes produced with a $^{82}$Se beam on beryllium targets,
whereas  open magenta circles and solid brown stars for isotopes obtained with a $^{76}$Ge beam~\cite{OT-PRC09}.
}
\end{minipage}\hspace{2pc}%
\begin{minipage}{18pc}
\vspace{-4.0cm}
\includegraphics[width=18.5pc]{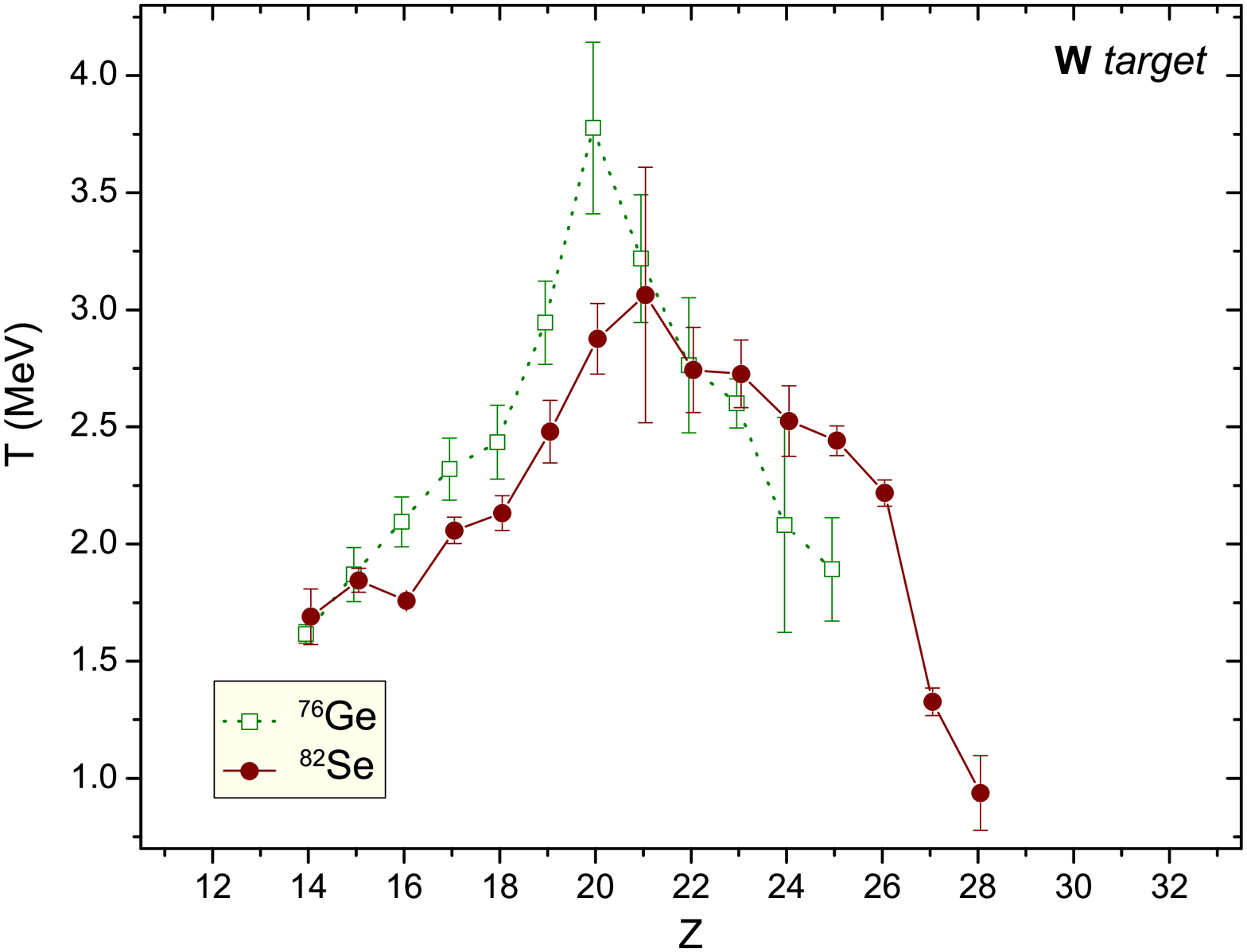}
\caption{(Color online) \label{Fig_TemperatureW} Similar to Fig.~\ref{Fig_TemperatureBe} for data with the tungsten target.}
\end{minipage}
\end{figure}

Fig.~\ref{Fig_Qg}
represents production cross sections measured with Be targets in this experiment,
where the abscissa, $Q^*_{g}$, is the smoothed
difference between the mass of the ground state of the projectile
and the observed fragment, where the masses were taken from Ref.~\cite{KTUY-PTP05}.
As in the previous experiment,  the heaviest isotopes of elements in the middle of the distribution
($Z=$~19, 20, 21, and 22) appear to break away from the
straight-line behavior.
The data were fitted by two lines with a floating connection point, and results are shown by lines in the figure.
The behavior of the slopes in the $Q^*_g$ figure are summarized in Fig.\ref{Fig_TemperatureBe} where
 the two individual fitted values of the inverse slope parameter, $T$, for products from Be targets are shown as a
function of atomic number.
The corresponding plot for the W target is shown in Fig.~\ref{Fig_TemperatureW}.
The inverse slopes of the cross sections from the previous experiment~\cite{OT-PRC09} with $^{76}$Ge beam are shown in these figures for comparison.

Based on these figures we find  that the general increase in
 $T$ for all of the heavy isotopes of elements $Z=$~19, 20, 21, and 22 observed with a $^{76}$Ge beam is reproduced by this experiment using the $^{82}$Se beam.

Small values of the inverse slope parameter $T$ for heavy neutron-rich isotopes of elements ($Z=$~27-33) in  Fig.\ref{Fig_TemperatureBe} can be explained by the fact that these isotopes were produced through transfer reactions (see the dashed lines in Fig.~\ref{Fig_CrossSection}, which show pick-up products), and therefore the well-known $Q_{gg}$-systematics  used for the two-body process may be more applicable.

\section{Summary\label{Summary}}

The present study of the fragmentation of a $^{82}$Se beam at
139~MeV/u provided evidence for the production of four previously
unobserved neutron-rich isotopes. The momentum
distributions and cross sections for a large number of neutron-rich
nuclei produced by the $^{82}$Se beam were measured by varying the
target thickness in a two-stage fragment separator using a narrow momentum selection.
The longitudinal momentum distributions of 122
neutron-rich isotopes of the elements with \protect{$11\le Z\le 32$}
were compared to models that describe the shape and centroid of
fragment momentum distributions.  New parameters for the
semiempirical momentum distribution models~\cite{DJM-PRC89,AG-PLB74} based on
the measured momenta were obtained.
The most neutron-rich nuclei of elements with $Z=19$ to 22 were
produced with an enhanced rate compared to the systematics
of the production cross sections from the $Q_g$ function.
This trend was previously reported for fragmentation with $^{76}$Ge beam
target~\cite{OT-PRC09}.

\ack
The authors would like to acknowledge the operations staff of the
NSCL for developing the intense $^{82}$Se beam necessary for this
study. This work was supported by the U.S.~National Science
Foundation under grant PHY-06-06007 and PHY-11-02511.


\section*{References}
\bibliography{82Se-NNC_v2}
\end{document}